# Effect of Ag(I), Ni(II), Zn(II) ions on Irradiated *Spirulina platensis*

E. Gelagutashvili, N.Bagdavadze, A.Gongadze, M.Gogebashvili, N.Ivanishvili

Iv. Javakhishvili Tbilisi State University
E. Andronikashvili Institute of Physics
0177, 6, Tamarashvili St.,
Tbilisi, Georgia
E.mail:eterige@gmail.com

## Abstract

Combined effects of $^{137}$Cs gamma irradiation and heavy metal ions Ni(II), Zn(II), Ag(I) on *Spirulina platensis* cells using UV-VIS spectrometry after three times irradiation and recultivation were discussed.
    It was shown, that possible use of gamma irradiation together with Ni(II) and Zn(II) ions does not change nature of interaction of these metal ions on *Spirulina platensis*. Whereas in contrast to the ions Ni (II) and Zn (II) for silver ions, an increase in intensity is observed in both the irradiated and non-irradiated states. The combined effects of ionizing radiation and other stressors such is silver ions for *Spirulina platensis* exhibit synergetic effects for C-phycocyanin as a stimulatory agent to raise the contents of it.

## Introduction

Cyanobacteria are the major biomass producers both in aquatic and terrestrial ecosystems [1]. In ancient times, cyanobacteria (blue-green algae) *Spirulina* was used as food by the Aztec civilization. *Spirulina* contains various components that are beneficial for health, such as proteins, vitamins, essential amino acids, minerals, γ-linoleic acid, glycolipids, sulfolipids, and phycobilins (phycocyanin, allophycocyanin, and phycoerythrin). Khan et al.[2] reported the main bioactive molecules and explained in detail their effects on health and human nutrition. Numerous studies have suggested that zeaxanthin and lutein are crucial for visual health. *Spirulina* can serve as a rich source of dietary zeaxanthin in humans [3].

    Some environmental stresses, including gamma irradiation and heavy metals, can influence on *Spirulina platensis*. In [4] were investigated the effect of gamma irradiation on the growth, biomass, pigment content and some metabolic activities of *A.platensis*. They reported that gamma irradiation had a stimulatory effect on its growth and cellular constituents. In our works [5,6] influence of 7.2 kGy gamma irradiation have been studied with optical and differential scanning microcalorimetry (DSC) methods on *Spirulina platensis* and also simultaneous effects of Cd(II), Pb(II) ions and γ-irradiation on stability of *Spirulina platensis* intact cells after gamma irradiation and without irradiation.

In this paper, discusses combined effects of $^{137}$Cs gamma irradiation and heavy metal ions Ni(II), Zn(II), Ag(I) on *S. platensis* cells using UV-VIS spectrometry after 3-times irradiation and recultivation.

## Materials and Methods

*Spirulina platensis* IPPAS B–256 strain was cultivated in a standard Zarrouk [7] alkaline water–salt medium at 34ºC, illumination ~5000lux, at constant mixing in batch cultures [8]. The cultivation of the *Spirulina platensis* cells was conducted for 14 days after irradiation. The growth was measured by optical density by monitoring of changes in absorption at wave length 560nm using the UV–Visible spectrometer Cintra 10e. The intact cells suspension of *Spirulina platensis* at pH 9.2 in the nutrition medium was used for scanning the absorption spectra from 400 to 800 nm. The concentration of *Spirulina platensis* was determined by the instrumental data [9,10]. *Spirulina platensis* suspension (100 ml) after 7 days of cultivation have been irradiated with 20kGy gamma radiation (Dose rate -1.1Gy during one minute) for 7 days using $^{137}$Cs as a gamma radiation source at the Applied Research Center, E.Andronikashvili Institute of Physics. Suspension after the irradiation (20 kGy) were filled up to 200ml with Zarrouk medium and they were recultivated. This irradiation and recultivation were repeated 3-told. The optical density was measured every day with 24h intervals. The concentration of different compounds were estimated at late exponential phase. The solutions of metal ions were prepared in deionizer water with appropriate amounts of inorganic salt ($AgNO_3$, $NiCl_2$, $ZnCl_2$).

## Results and Discussions

Figure 1 shows the growth kinetic curves of the cyanobacterium *Spirulina platensis*: control, and irradiated control after 3 times irradiation (every case irradiation dose 20 kGy) and recultivation. The curves present dependence of the optical density at 560 nm on cultivation time As it is seen from the same figure 1, the *Spirulina platensis* control have been monitored during 1 week of incubation, and the irradiated suspension have been monitored for 2 weeks. After 2 weeks, the optical density of the irradiated suspension after 3-told irradiation and recultivation is 87% of cultivation result of the control.

Metal effect by integral irradiated and recultivated three times cells of blue-green microalgae *Spirulina platensis* was studied as a function of metal concentration (pH 9.3). Fig. 2 shows the absorption spectra of control and the same control after three times irradiation (every case irradiation dose 20 kGy) and recultivation of *Spirulina platensis*. The peak at 681 nm is due to the absorption of Chl a peak. At 621 nm is due to the absorption of phycocyanin (PC). A peak at 440 nm is due to soret band of Chl a [11]. Comparison of absorption spectra (Figure 2) shows that absorption for irradiated cells decreases slightly.

In fig.3 are shown effect of Ni(II), Zn(II), Ag(I) ions on the absorption of the intact cells after irradiation and recultivation of *Spirulina platensis* and without irradiation.It is seen from fig.3, that with increasing metal concentrations absorption intensity decreased for Ni(II), Zn(II) metal ions and increased for Ag(I) ions. Such effects were observed both in irradiation and nonirradiation cases. As can be seen from this figure, the absorption processes was relatively fast in the small concentrations for Ni(II) and Zn(II) ions and then became slow.

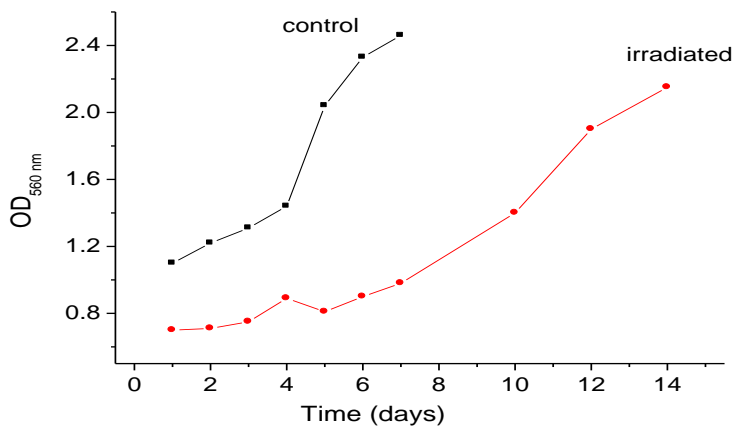

Fig.1 Growth kinetic curves of the cyanobacterium *Spirulina platensis*: control and irradiated control after three times irradiation (every case irradiation dose 20 kGy) and recultivation.

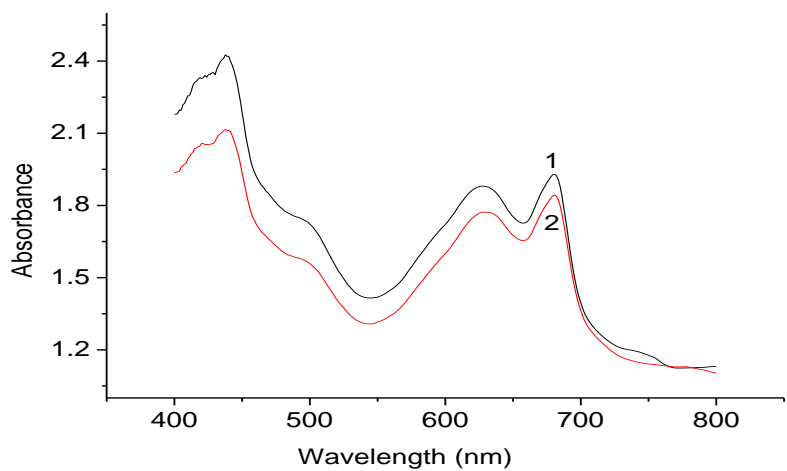

Fig.2. The absorption spectra of the cells of *Spirulina platensis* recorded after cultivation
    1 – Control after incubation 7 days
    2 – Suspension after three times γ-irradiation and after incubation for 14 days

By us were also investigated influence of the same metal ions on the same cellular components of *Spirulina platensis* without irradiation and the same irradiation dose. Effect of metal ions on the absorption intensity maximums for wavelengths 440nm, 621 nm and 681nm are shown in fig.4. At 440 nm with increasing Ag(I) concentration the intensity of absorption increases, almost does not change for Ni(II) ions along this wavelength both cases in irradiation and nonirradiation cases. Increasing the concentration for Zn (II) ions causes a decrease in absorption intensity in the case nonirradiation, but almost does not change for irradiation case.

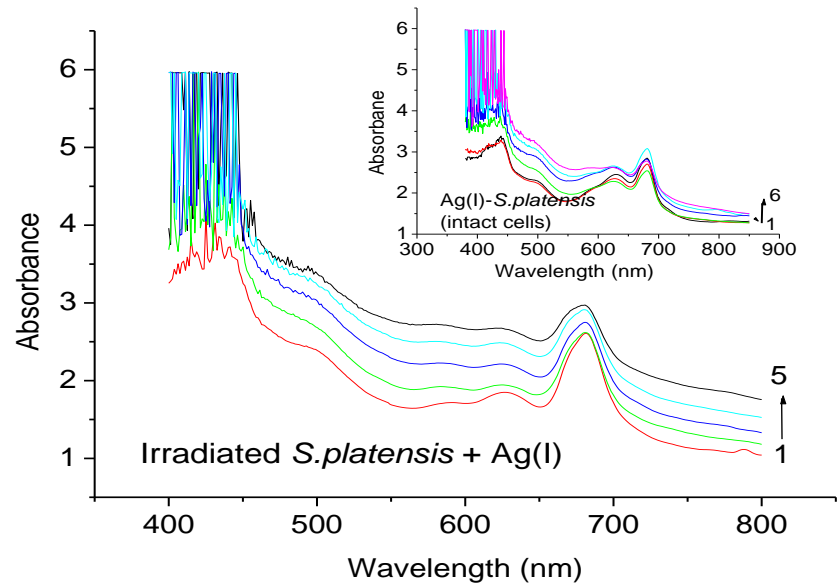

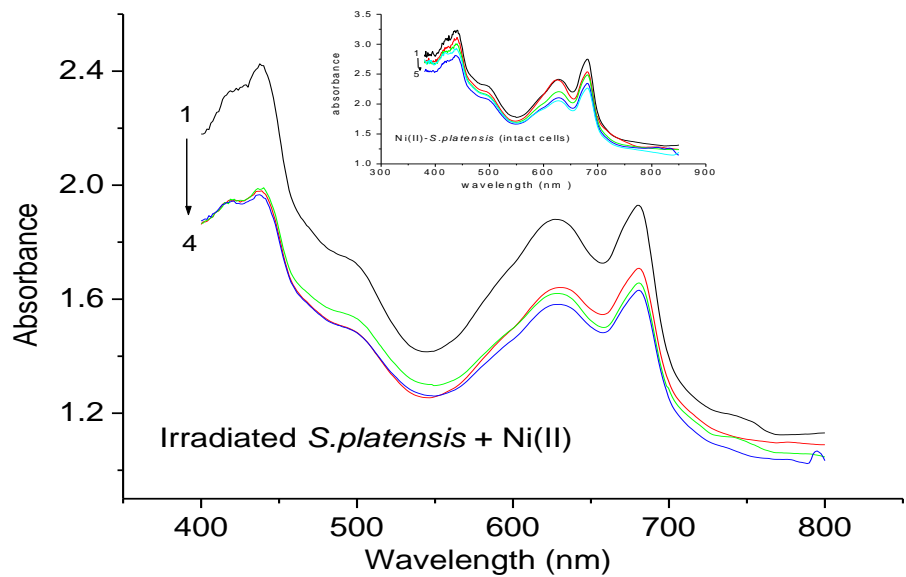

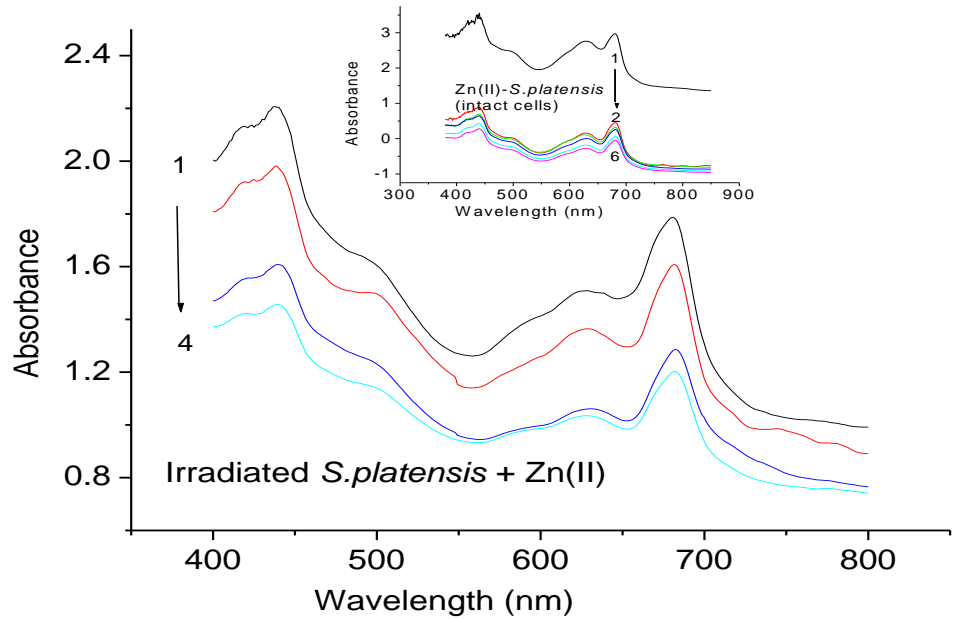

Fig.3.Effect of Zn(II), Ni(II) and Ag(I) ions on the absorption of irradiated suspension of *Spirulina platensis* after three times irradiatio and recultivation: 1→4 [Zn(II)]=0 ÷3 mM; 1→4 [Ni(II)]= 0 ÷3 mM; 1→5 [Ag(I)]=0 ÷4 mM; insert: Effect of Zn(II), Ni(II) and Ag(I) ions on the absorption of the intact cells of *Spirulina platensis*. 1→6 [Zn(II)]=0 ÷5 mM; 1→5 [Ni(II)]= 0 ÷4 mM; 1→6 [Ag(I)]=0 ÷5 mM;

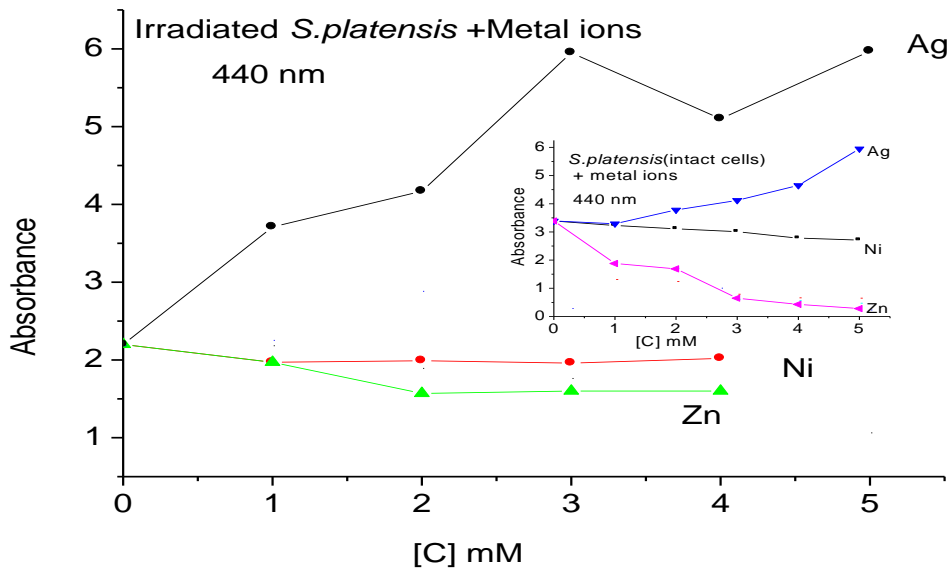

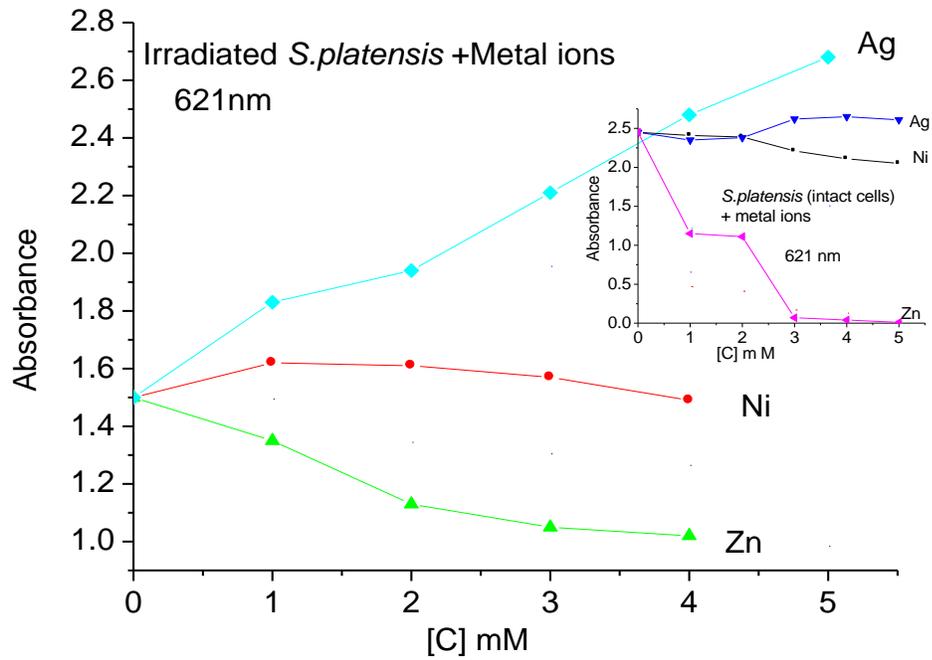

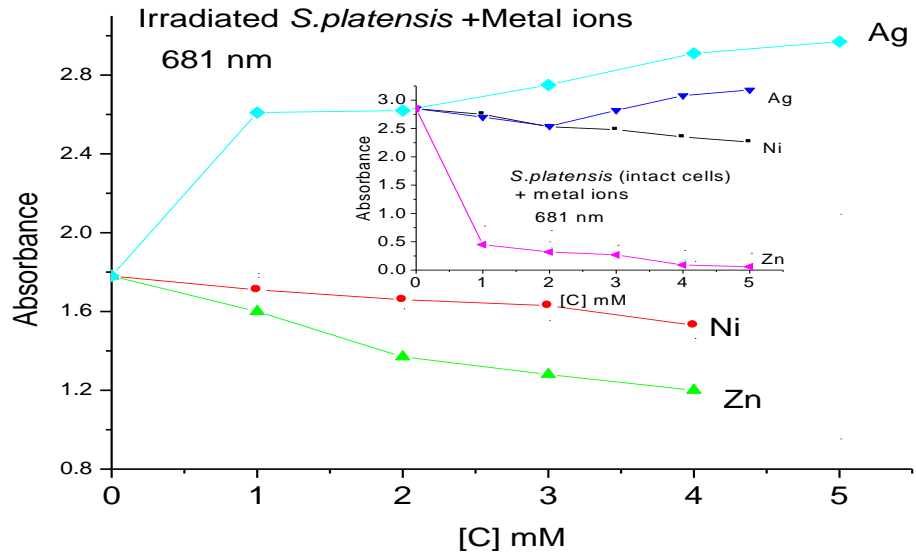

Fig.4. Changes in the absorption intensity of *Spirulina platensis* at 621 nm (phycocyanin), at 440 nm (the soret band of Chl a), at 681 nm (the Chl a) under the influence various metal ions after irradiation. insert: Effect of the same metal ions on the same cellular components of *Spirulina platensis* without irradiation

As for the change in absorption intensity at 621 nm, which is the peak of absorption of the major protein-phycocyanin of *Spirulina platensis* (figure 4), as the concentration of silver ions increases, the absorption intensity very efficiently increases in the case of irradiation ,but almost does not change without irradiation. But unlike silver ions, an increase in the concentration of zinc causes the most noticeable decrease in the intensity of absorption both case in irradiation and nonirradiation case. For nickel such changes does not observed in both case. Similar results were observed in absorption intensity for Ag(I) , Ni(II) and Zn(II) ions at 681 nm, which is the absorption peak of the Ch a of *Spirulina platensis*. In particular, that as the concentration of silver ions increases, the absorption intensity increases very efficiently in the case of irradiation and without irradiation. An increase in the concentration of zinc causes the most noticeable decrease in the intensity of absorption intensity without irradiation, than in the irradiation case. At 681 nm for nickel ions were observed, changes of the same nature as in the case of 621 nm.

Thus, this study show that possible use of gamma irradiation together with Ni(II) and Zn(II) ions does not change nature of interaction of these metal ions on *Spirulina platensis*. Whereas in contrast to the ions Ni (II) and Zn (II) for silver ions, an increase in intensity is observed in both the irradiated and non-irradiated states. The combined effects of ionizing radiation and other stressors such is silver ions for *Spirulina platensis* exhibit synergetic effects for C-phycocyanin as a stimulatory agent to raise the contents of it.